# Pulse profiles of millisecond pulsars

Ray Jayawardhana & Jonathan E. Grindlay

*Harvard-Smithsonian Center for Astrophysics*

## 1. Introduction

We present a comparison between the average radio pulse profiles of millisecond pulsars (MSPs) in the field and in globular clusters. Our sample consists of 20 field MSPs and 25 cluster MSPs for which observations exist at 400 - 600 MHz.

## 2. Presence of an interpulse

We find that 6 of the 20 field MSPs, or about 30%, have a comparable interpulse at a phase offset of 180 ±30 degrees. None of the cluster objects shows this feature. Here we define a "comparable interpulse" as one whose intensity is at least 10% of the primary pulse. This lower limit is well above the noise level of all the profiles in our sample. While the cluster MSPs have much more poorly resolved profiles at present than field MSPs, it is unlikely that a strong interpulse near 180 degrees could be "hidden" within the primary peak.

Table 1. Millisecond pulsars with an interpulse

| Name | Fractional intensity | Phase offset (degrees) |
| --- | --- | --- |
| PSR 1937+21 | 0.56 ±0.05 | 175 ± 4 |
| PSR 1957+20 | 0.34 ±0.05 | 165 ± 4 |
| PSR 1855+09 | 0.30 ±0.05 | 200 ± 4 |
| PSR 2322+2057 | 0.13 ±0.05 | 170 ± 4 |
| PSR 1012+5307 | 0.29 ±0.05 | 192 ± 4 |
| PSR 1913+16 | 0.59 ±0.05 | 191 ± 4 |

## 3. Pulse Width - Period relation

We have also measured the full width at half-intensity of the primary peak in degrees for all MSPs in our sample. A least-square fit to the cluster MSPs gives a simple power-law relation between pulse width and period, with an index of -0.55 ±0.06 (Figure 1). Interestingly, this dependence roughly agrees with that predicted by the dipole polar cap models for canonical pulsars. Field MSPs, on the other hand, show a much larger scatter in this diagram. Note also that the six field MSPs with an interpulse do not form an obvious distribution.

Given the poor time resolution of cluster MSPs, it is not easy to identify their primary peaks as Single, Double, or Multiple, as Backer (1984, Ap & Astron., 5, 187) did for a few field MSPs.

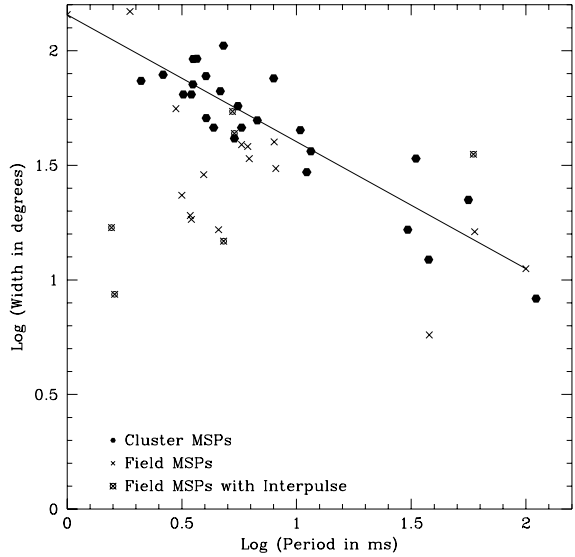

Figure 1.   Pulse width v. Period

## 4.   Pulse width distribution

A Kolomogorov-Smirnoff test of the cumulative width distributions of field and cluster MSPs gives a 77% probability that they are drawn from two different populations. For cluster and field MSPs with periods less than 5ms, the equivalent probability goes up to 83 %.

However, cluster MSPs at present have much more poorly resolved profiles than field MSPs. Thus it is possible that narrower peaks would emerge in cluster MSP profiles with improved time resolution.

## 5.   Discussion

The differences in the pulse properties of field and cluster MSPs are suggestive of two different populations, possibly with different origins (see, for example, Chen & Ruderman, 1993, ApJ, 408, 179). A larger sample of MSP profiles with higher time resolution, particularly for those in clusters, as well as polarization studies may help test this possibility.

**Acknowledgments.**   It is a pleasure to thank S. Anderson, Z. Arzoumanian, M. Bailes, F. Camilo, W. Deich, S. Lundgren, R. Manchester, J. Navarro, P. Ray, and S. Thorsett for their help in obtaining the pulse profiles. We also gratefully acknowledge the authors of the many papers which provided data for this study. Unfortunately, space limitations do not allow us to reference them here.